\documentstyle[editedbook,epsfig,psfig,epsf]{mq}
\begin{opening}
\title{State transition between high-soft and low-soft states observed in
GRS~1915$+$105}
\author{S. Naik$^1$, A. R. Rao$^1$, and Sandip K. Chakrabarti$^2$}
\institute{$^1$ Tata Institute of Fundamental Research, Mumbai, 400 005, India.
\\
$^2$ S.N. Bose National Center for Basic Sciences, Calcutta, 700 091, India}
\end{opening}
\runningtitle{State transition in GRS~1915$+$105}
\runningauthor{Naik S., et al.}

\begin{document}
\vspace{-0.5cm}
\begin{abstract}
{\small We present the results of a detailed analysis of RXTE observations
which show an unusual state transition between high-soft and low-soft states 
in the microquasar GRS~1915+105. The RXTE pointed observations reveal that 
these events appeared as a series of quasi-regular dips when hard X-ray and
radio flux were very low. The X-ray light curve and colour-colour diagram
of the source during these observations are found to be different from any
reported so far. The X-ray spectral and timing properties of the source during 
these dips are distinctly different from those seen during the other various 
dips and low intensity states seen in this source. There is, however, a 
remarkable similarity in the properties during the dip and non-dip regions 
in these set of observations. This indicates that the basic accretion disk 
structure is similar, but differ only in intensity. We invoke a model, to 
explain these observations, in which the viscosity is very close to a 
critical viscosity and the shock wave is weak or absent.}
\end{abstract}

\vspace{-0.5cm}
\section{Introduction}

In spite of the bewildering types of variability classes seen in GRS~1915$+$105,
the spectral and temporal characteristics at any given time can broadly be 
classified into three distinct spectral states: a low-hard state with invisible 
inner accretion disk (C), a high-soft state with visible inner accretion disk 
(B), and a low-soft state with spectrum similar to the high-soft state and with 
much lower intensity (A). All the variability classes are understood as 
transition among these three basic states.  In GRS~1915$+$105, the low-soft 
state appears briefly during the rapid state transition between states C and 
B as well as during the soft dips seen during the variability classes $\beta$ 
and $\theta$ (Belloni et al. 2000). On rare occasions, long stretches of state 
A are also seen in this source (the variability class $\phi$).
Though transition from low-hard to high-soft states are 
seen in many Galactic black hole candidate sources, a transition between two 
different intensity states with similar physical parameters of the accretion 
disk was not observed in GRS~1915$+$105 or in any other black hole binaries.
We have tried to explain the observed peculiar state transition which is
seen for the first time in GRS~1915$+$105 on the basis of the presence of
an accretion disk with critical viscosity which causes appearance and
disappearance of sub-Keplerian flows out of Keplerian matter.

\vspace{-0.5cm}
\section{Observations and Analysis}

We have made a detailed examination of all the publicly available
RXTE pointed observations on GRS~1915$+$105 in conjunction with the 
RXTE/ASM data. During these investigations we found a new variability 
class of observations which was occurring during two time intervals, 
1999 April 23$-$May 08 and 1999 August 23$-$September 11, 
respectively. This new class (class $\omega$; Klein-Wolt et al. 2002) 

\begin{minipage}{70mm}
was observed in a total of 16 pointed RXTE observations and is 
characterized by a series of dips of duration of 20$-$95 s and repetition 
rate of 200$-$600 s. We show in Figure 1 the X-ray light curve and the 
colour-colour diagram for a few of the variability classes which show dip 
like structure or prolonged low intensity states. We have defined the 
hardness ratio HR1 as the ratio between the count rate in the energy range 
5$-$13 keV to that in 2$-$5 keV and HR2 as the ratio of the count rates 
in the energy range 13$-$60 keV to that in 2$-$13 keV. From the figure, 
it is observed that the properties of the source during the new class is 
distinctly different from state C. The most remarkable feature in the new 
class is the low value of HR1, in fact the lowest value seen during any 
variability class observed in GRS~1915+105. 

To study the timing properties of the source, we have generated
the power density spectrum (PDS) in 2$-$13.1 keV energy band for the dip
and non-dip regions of all 16 selected RXTE pointed observations. 
It is found that the low frequency narrow QPOs are absent in the PDS of 
dip and non-dip regions of all the observations and the PDS is a power-law with
with indices of $-$1.5 to $-$1.9 and $-$0.8 to $-$1.9, during the non-dip 
and dip regions, respectively.
\end{minipage}
\hfil\hspace{\fill}
\begin{minipage}{55mm}
\hspace{1.5cm}
\includegraphics[width=54mm]{naiks_1.ps}
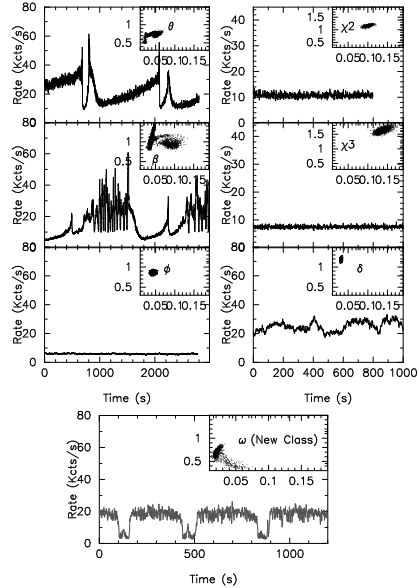
{\small Fig. 1. The X-ray light curves (2$-$60 keV energy range) and
colour-colour diagram (HR1 vs HR2) of GRS~1915$+$105 for observations of
classes $\beta$, $\theta$, $\phi$, $\chi$2, $\chi$3, and $\delta$ 
are shown along with the observation of class $\omega$. The insets 
in each figure shows the colour-colour diagram, HR1 in the Y-axes 
and HR2 in the X-axes (see text).\\
\\}
\end{minipage}

We have attempted a wide band X-ray spectroscopy of the dip and 
non-dip regions of the RXTE observation on 1999 August 23. The data
for the dip region were selected when the source count rate was 
$\leq$ 3000 counts s$^{-1}$ (for 2 PCUs) in 2$-$60 keV energy range and
non-dip region when the count rate was $\geq$ 5000 counts s$^{-1}$.
Standard procedures for data selection, background estimation and response
matrix generation have been applied. 3$-$50 keV energy range PCA data and 
25$-$180 keV energy range HEXTE data are used for spectral fitting.
The value of absorption column density (N$_H$) was kept fixed at 6 $\times$
10$^{22}$ cm$^{-2}$. The three component spectral model with a disk-blackbody,
a power-law, and a thermal-Compton spectrum as model components fits the data
during both the dip and non-dip regions with reduced $\chi^2$ of 0.68 (66 dof)
and 0.69 (76 dof) respectively.
We have shown, in Figure 2, the energy spectra and the fitted
model during two different intensity states. From the figure, it is 
observed that the dip and non-dip spectra are dominated by the thermal 
component. We have calculated the luminosity of the source in 3$-$60 keV 
energy band to be $\sim$ 7.8 $\times$ 10$^{38}$ ergs s$^{-1}$ and 2.2 $\times$
10$^{38}$ ergs s$^{-1}$ for non-dip and dip regions respectively.
The calculated values of the inner disk radius are 42$\pm$5 km during 
the non-dip and 25$\pm$6 km during the dip regions 
(assuming $d$ = 12.5 kpc and $i$ = 70$^{\circ}$). The temperature of the 
inner accretion disk during the dip and non-dip regions are found to be 
\begin{minipage}{70mm} 
1.5$_{-0.1}^{+0.1}$ keV and 1.7$_{-0.1}^{+0.1}$ keV respectively. From 
this analysis, we conclude that the observed transition between the 
different intensity states are associated with the change in temperature 
of the inner accretion disk without any significant change in the radius. 

\vspace{-3mm}
\section{Discussion}
We now discuss a possible scenario to understand this new and exciting behavior 
in terms of TCAF model of Chakrabarti \& Titarchuk (1995). A disk with 
completely free (Keplerian) disk and 
(sub-Keplerian) halo accretion rates, does exhibit multiplicity in spectral 
index or multiplicity in disk accretion rate when the spectral index is 
similar. In the observation described here, there is 
no evidence for a large  variation of the spectral index. This signifies that
though there is not enough time for a change in the total rate,
individually, Keplerian and sub-Keplerian rates may have been modified.
This is possible if the viscosity parameter is very close to the critical
value ($\alpha \sim \alpha_c \sim 0.015$; Chakrabarti 
1996). When $\alpha$ of the entire disk is well above $\alpha_c$, 
the entire disk is pretty much
\end{minipage}
\hfil\hspace{\fill}
\begin{minipage}{60mm}
\includegraphics[width=51mm]{naiks_2.ps}
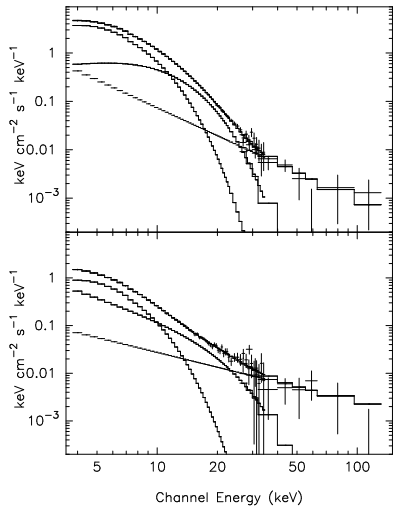
{\small Fig.~2.~The observed count rate spectrum of GRS~1915+105 
during non-dip and dip regions of the new class obtained from RXTE/PCA 
and HEXTE are shown in upper and lower panels respectively. A best-fit 
model consisting of a disk blackbody, a power-law, and a thermal 
Compton spectrum is shown as histogram with the data.}
\end{minipage}
Keplerian, except very close to the black hole ($r<3r_g$). 
Similarly when $\alpha$ of the entire flow is well below $\alpha_c$, the 
flow is sub-Keplerian with a possible standing or oscillating shock wave.
Since viscosity in a disk can change in a very small time scale 
(convective/turbulent time-scale in the vertical direction) it is not 
unlikely that the viscosity near the Keplerian-disk surface is very close 
to the critical value during the time when this new class is exhibited. 
Temperature and intensity of the radiation from the Keplerian disk drops 
to the point that a thin outflow develops from the sub-Keplerian flow. This 
is the low intensity state. This wind is cooled down and is fallen back on 
the Keplerian disk, increasing the Keplerian rate, intensity and viscosity, 
thereby cutting off the wind and bringing the flow to the high intensity 
state again. As the disk is cooler, no shock oscillation, and therefore
no QPO is present. One of the tests that our model is perhaps correct is that
because of super-critical viscosity this new class ($\omega$) should be in
association with other soft classes ($\delta, \ \phi $ and $\gamma$) as is
indeed the case. Only when the net accretion rate or viscosity goes down,
the spectrum may become partly harder and $\theta$ and $\beta$
states could occur. If viscosity goes up, then the state goes to
$\delta$, $\gamma$ or $\phi$ and the disk is basically Keplerian
in these cases.

\vspace{-0.4cm}
\section*{References}
Belloni, T., Klein-Wolt, et al. 2000, A\&A, 355, 271\\
Chakrabarti, S. K. 1996, ApJ, 464, 664\\
Chakrabarti, S.K. \& Titarchuk, L.G. 1995, ApJ, 455, 623\\
Klein-Wolt, M., Fender, R. P., et al. 2002, MNRAS, 331, 745
\end{document}